\documentclass[12pt,preprint]{aastex}
\begin{document}
\title{A New Perspective on the Large-Scale Tidal Effect on the Galaxy Luminosity and Morphology}
\author{Jounghun Lee}
\affil{Astronomy Program, Department of Physics and Astronomy, Seoul National University, 
Seoul 08826, Republic of Korea} 
\email{jounghun@astro.snu.ac.kr}
\begin{abstract}
We study the mean tidal coherence of galaxy environments as a function of intrinsic luminosity determined by the absolute 
$r$-band magnitude.  The tidal coherence of a galaxy environment is estimated as the cosine of the angle between two minor eigenvectors 
of the tidal field smoothed at the scales of $2$ and $30\,h^{-1}$Mpc, respectively, centered on each of the local galaxies from the Sloan 
Digital Sky Data Release 10. Creating four luminosity-selected samples of the Sloan galaxies, we control them to have identical density 
distributions, in order to nullify the dominant effect of the local density.  It is found that the samples containing more luminous 
wall and field galaxies yield lower mean values of the tidal coherence, which trend turns out to be robust against the variation 
of the smoothing scales. At fixed morphology, the same trend is found for the late-type spiral and lenticular galaxies in both of the 
field and wall environments. 
The early-type spiral field galaxies show no significant dependence on the tidal coherence, while both of the least and most luminous 
elliptical wall galaxies are found to dwell in the regions with highest tidal coherence. 
\end{abstract}

\section{Introduction}\label{sec:intro}

Ever since  \citet{dre80} observationally discovered the dependence of the morphological types of the cluster galaxies on the local 
densities, numerous observational studies have confirmed the existence of the correlations between the local environmental densities and 
various physical properties of the galaxies, which include the luminosity, morphological shape, color, star formation rate, metallicity, and so 
on \citep[e.g.,][]{hay-etal84,PG84,san-etal92,bal-etal98,got-etal03,hog-etal03,hog-etal04,kau-etal04,tan-etal04,bla-etal05,zeh-etal05,KR05,
yee-etal05,ber-etal06,sol-etal06,cap-etal07,den-etal07,omi-etal08,wel08,bam-etal09,ide-etal09,ski-etal09,tas-etal09,wei-etal09}. 
To address the critical issue of what contributed most to establishing this environmental dependence of the galaxy properties, 
much theoretical and numerical endeavor has been made
\citep[e.g., see][and references therein]{avi-etal05,AS05,coo05,CZ05,har-etal06,lee06,per-etal06,mau-etal07,BB07,KO08,cro-etal09,
kim-etal09,par-etal09,VR13,TC14,met-etal15,jos-etal17,lee-etal17,gup-etal18}. 

In fact, the observed environmental dependence of the galaxy properties is not well accommodated by the simplest excursion set theory 
which explains that the formation and evolution of a dark matter halo is conditioned solely on its mass, regardless of the local 
environmental density \citep{GG72,PS74,bon-etal91}.  
Nonetheless, it has been generally believed that the observed environmental dependence of the galaxy properties may be 
fitted into the standard structure formation scenario based on the $\Lambda$CDM ($\Lambda$ and cold dark matter) cosmology 
if the effects of gravitational merging and other astrophysical processes on the formation and evolution of the galaxies are 
properly taken into account since those processes depend on the environmental densities 
\citep[e.g.,][]{avi-etal05,mat-etal07,mau-etal07,tas-etal09,FM10,jos-etal17}

In the conventional approaches,  the local density was regarded as the primary feature that plays the most decisive role in 
establishing the link between the physical properties of galaxies and their environments.  The subject has been gradually 
extended and improved to embrace the secondary environmental features other than the local density, in the hope that their 
independent additional effects on molding the galaxy properties, if existent and detected, might allow us to find a missing puzzle 
in our understanding of the galaxy formation. 
Especially,  since the cosmic web phenomenon was attributed to the large-scale coherence of the tidal shear field \citep{web96}, 
the possible effect of the large-scale tidal shear on the formation and evolution of the galaxies embedded in the cosmic web has risen 
as a topical issue. 
\citet{PB06} showed that the more intrinsically luminous galaxies are observed to exhibit lower degree of anisotropy in their spatial 
distributions. \citet{LL08} found observationally that at constant local density the elliptical and lenticular galaxies preferentially reside in the 
regions with weaker tidal shears (quantified in terms of regional ellipticity) than the spiral counterparts
 \citep[see also][]{nuz-etal14,pou-etal17}. 
\citet{lai-etal18} found both observationally and numerically that the locations of the more massive galaxies are closer to the densest 
sections of the filaments. 

However, the importance of the large-scale tidal effect on the galaxy properties was challenged by several authors who pointed out that the 
detected signals were either too weak to be physically significant or merely reflection of strong correlations of the tidal shears with the local 
densities.  \citet{yan-etal13} detected observationally a signal of correlation between the environmental shear and the galaxy morphology, 
which turned out to be in fact quite consistent with the result of \citet{LL08}.  Instead of confirming the previous work, however, 
\citet{yan-etal13} refuted it with the claim that the detected signal of correlation bears only marginal physical significance as it seemed not induced 
by the independent tidal effect \citep[see also][]{zhe-etal17}.
\citet{par-etal18} examined the effect of the tidal shear on the redshift-space two-point density correlations of the group galaxies by 
analyzing observational data and concluded that including only the mass effect on the galaxy evolution suffices to explain the apparent 
correlation between the galaxy clustering and the tidal environments \citep[see also][]{ala-etal18}. 
Very recently, \citet{goh-etal18} studied the properties of the galactic halos located in different large-scale structures from a high-
resolution simulation and claimed non-existence of any evidence for the variation of the halo properties with the geometrical shapes of 
the web elements at fixed density.

Yet, it is premature to conclude that the large-scale tidal environment plays no independent role in the formation and 
evolution of the galaxies. The previous works which explored the large-scale tidal effect on the galaxy evolution, whether they 
found evidences or counter evidences, implicitly assumed that the variation of the galaxy properties would be induced by the 
differences in the strength of the tidal shear (or equivalently degree of anisotropy or geometry of the cosmic web), which was 
quantified in terms of the {\it eigenvalues} of the tidal fields. 

The seminal work of \citet{PS17} based on Shannon's information theory \citep{sha48}, however, gives an enlightening hint 
that what may affects most the galaxy properties is not the strength but the coherence of the environmental effects. 
What they studied is the conditional mutual information between the morphological shapes of the galaxies and the large-scale 
environments smoothed on various mass scales. Showing that the conditional mutual information decreases with the increasing smoothing 
scale but remains non-negligible even on the scale of $30\,h^{-1}$Mpc, \citet{PS17} suggested that the environmental dependence of the 
properties of the galaxies should be an outcome of the mutual correlations of the density environments at largely separated scales. 

Motivated by the idea of \citet{PS17} and extending their logic to the tidal environments, we attempt here to investigate if and how 
the properties of the galaxies observed in the universe show any unique dependence on the coherence of the tidal field over large-scales 
quantified in terms of the {\it eigenvectors} of the tidal fields.  
For this investigation, we will consider only the absolute $r$-band magnitudes (i.e., luminosity) and morphological types as the 
main intrinsic properties of the galaxies, given the result of \citet{par-etal07} that no other property shows an independent link to the 
environment at fixed luminosity and morphological type.
The organization of this paper is as follows. In Section \ref{sec:data} we will briefly describe how to create controlled samples of the 
galaxies from observational data and how to determine the tidal coherence at each location of the sample galaxies. 
In Section \ref{sec:eecor_lum}, we will present how the absolute $r$-band magnitudes of the galaxies depend on the coherence of the tidal 
fields over large-scales. In Section \ref{sec:eecor_type}, we will present how the dependence of the galaxy luminosity on the tidal 
coherence changes with galaxy morphology. The summary and discussion of the main results will be presented in Section \ref{sec:con}.

\section{Description of the Observational Data}\label{sec:data}

\citet{wan-etal09} developed a novel method (called the "halo tracing" algorithm) with which the density contrast, $\delta({\bf r})$, 
and peculiar velocity, ${\bf v}({\bf r})$, fields can be statistically constructed from a finite set of group-sized dark halos with masses larger 
than a certain critical value.
Using the groups of the local galaxies identified by the Sloan Digital Sky Survey  in the redshift range of $0.01\le z\le 0.4$ as proxies of 
dark halos, \citet{wan-etal12} succeeded in making the first observational application of this method and demonstrated its efficiency. 
Fitting the Sloan survey volume, $V_{\rm S}$, to a rectangular box consisting of $494\times 892\times 499$ grids, \citet{wan-etal12} 
calculated the realizations of $\delta({\bf r}_{i})$, and ${\bf v}({\bf r}_{i})$ on each grid center ${\bf r}_{i}$ and asserted the reliability of 
the realizations in the inner subvolume of $\sim 0.66V_{\rm S}$ upon the condition of the smoothing scale, 
$R_{f}\ge 2\,h^{-1}$Mpc. 

Utilizing the data of of $\delta({\bf r}_{i})$ publicly released by \citet{wan-etal12}, \citet{LC15} linearly constructed the tidal shear field, 
${\bf T}({\bf r})$, in the innermost cubic box of volume $V_{\rm I}=180.1^{3}\,h^{-3}$Mpc$^{3}$ that includes $256^{3}$ grids 
in the ranges of $-182.28\le x/(h^{-1}\,{\rm Mpc})\le -1.49$, $-90.05\le y/(h^{-1}\,{\rm Mpc})\le 90.74$, 
$-22.73\le z/(h^{-1}\,{\rm Mpc})\le 158.07$, where $(x,\ y,\ z)$ represents a position vector in the equatorial Cartesian coordinate system.  
Performing a Fourier transformation of $\delta({\bf r})$ to find its Fourier amplitude, $\tilde{\delta}({\bf k})$, they computed the Fourier 
amplitudes of the tidal shear tensors, $\tilde{\bf T}({\bf k})$, as
\begin{equation}
\label{eqn:Tk}
\tilde{T}_{ij}({\bf k})\equiv \frac{k_{i}k_{j}}{k^{2}}\tilde{\delta}({\bf k})\exp\left(-\frac{k^{2}R^{2}_{f}}{2}\right)\, ,
\end{equation}
where ${\bf k}=(k_{i})$ is the wave vector with $k=\vert{\bf k}\vert$ in the Fourier space. An inverse Fourier transformation of 
$\tilde{\bf T}({\bf k})$ yielded the tidal field, ${\bf T}({\bf r};R_{f})$, smoothed with a Gaussian filter of scale radius $R_{f}$, 
which we are going to use throughout the current analysis.

We also utilize the flux-limited galaxy catalog compiled by \citet{tem-etal14} from the Tenth Data Release of the Sloan Digital Sky Survey 
\citep{sdssdr10}.  The catalog informs us of various spectroscopic properties of the Sloan galaxies such as their equatorial Cartesian 
coordinates, $(x_{g},\ y_{g},\ z_{g})$, absolute $r$-band magnitude, $M_{r}$, and morphological classifications 
given in terms of the probabilities of being classified as elliptical, lenticular, early-type spiral and late-type spiral galaxies, 
$P({\rm E}),\ P({\rm S0}),\ P({\rm Sab})$ and $P({\rm Scd})$, respectively.   
Using information on $(x_{g},\ y_{g},\ z_{g})$, we select those as targets from the catalog, which reside within  $V_{\rm I}$,  
ending up having a total of the $203,040$ targets. 
The position, $(x_{g},\ y_{g},\ z_{g})$, of each target is also used to determine its nearest grid (called the galaxy grid, hereafter). 

At each galaxy grid, we find an eigenvector, $\hat{\bf e}$, of ${\bf T}({\bf r})$ that is a solution to the equation of 
${\bf T}\hat{\bf e} = \lambda\hat{\bf e}$ where $\lambda$ denotes the corresponding eigenvalue. Since ${\bf T}=(T_{ij})$ at each 
galaxy grid  is a real $3\times 3$ symmetric matrix, the above equation has three solutions. 
The minor eigenvector corresponds to the smallest eigenvalue (say, $\lambda$), among the three, whose direction is parallel to the 
direction along which the surrounding matter is least compressed or equivalently most stretched. That is, this is the direction along 
which the tidal shear fields are coherent over the largest scales. 

The minor eigenvector is obviously a function of the smoothing scale, i.e., ${\bf e}={\bf e}(R_{f})$. 
Smoothing the tidal field on two different scales: $R_{f1}=2\,h^{-1}$Mpc  and $R_{f2}=30\,h^{-1}$Mpc,   
we find two minor eigenvectors ${\bf e}(R_{f1})$ and ${\bf e}(R_{f2})$, and compute the tidal coherence defined as 
$\vert{\bf e}(R_{f1})\cdot{\bf e}(R_{f2})\vert$ at each galaxy grid. 
It is worth explaining the reason for our choice of these specific values of $R_{f1}$ and $R_{f2}$. 
The former choice, $R_{f1}=2\,h^{-1}$Mpc, is made to represent the nonlinear scale on which the local environmental 
exhibits the strongest correlation with the galaxy properties  \citep[e.g.,][]{kau-etal04,yan-etal13}. 
Whereas, our choice of latter, $R_{f2}=30\,h^{-1}$Mpc, is made in light of the result of \citet{PS17} that the maximum scale up to 
which the mutual information between the galaxy morphology and the large-scale environments retains its non-negligible level. 

\section{Dependence of the Galaxy Luminosity on the Tidal Coherence}\label{sec:eecor_lum}

To investigate if and how strongly the galaxy luminosity depends on the tidal coherence, we divide the selected target 
galaxies according to their absolute $r$-band magnitudes, $M_{r}$, into four samples, say {\cal I}, {\cal II}, {\cal III} and {\cal IV}  
(See Table \ref{tab:mr}).  The sample {\cal I} contains the most luminous galaxies while the dimmest galaxies belong to the 
sample {\cal IV}. For each luminosity-selected sample, we classify the member galaxies into the field and wall category, 
which are defined to satisfy the conditions of $\delta(R_{f1})<1$ and $\delta(R_{f1})\ge 1$, respectively. 
The wall galaxies grow mainly through merging and infall of satellites, while the diffusive accretion is largely responsible for the 
growth of the field counterparts \citep{FM10}. To take into account this difference between the wall and the field galaxies 
in the mechanism that drives their evolutions, we will treat the wall and field galaxies separately in our investigation of the effect of 
the tidal coherence on their luminosities.

We are going to calculate the mean tidal coherence, $\langle\vert{\bf e}(R_{f1})\cdot{\bf e}(R_{f2})\vert\rangle$, for each 
luminosity-selected sample and then see whether or not a different sample yields a different value of tidal coherence. 
This investigation, however, must be preceded by an important task, which is nothing but to nullify the effect of the local density. 
The local environments where the member galaxies of each sample are embedded may well have different densities. Given that the 
physical properties of the galaxies depend quite strongly on the local density and that the tidal field by definition is correlated with the local 
density, the difference in the local density between the samples could cause a false positive signal of correlation between the tidal 
coherence and the galaxy properties. 
In the previous works which dealt with the tidal effect on the galaxy properties, this task was carried out by reducing the size of the 
galaxy sample through constraining the local density to a narrow range \citep[e.g.,][]{LL08}. Although this approach allowed one to 
minimize the effect of the local density in the most conservative way, it inevitably resulted in a low signal-to-noise ratio due to the 
reduced sample size. 

A more efficient way to nullify the effect of the local density without substantially reducing the sample size is to 
control the different samples to have identical density distributions.  
If the controlled samples show any difference in the values of tidal coherence, then this difference should not be ascribed to the effect of 
the local density since the four controlled samples have identical density distributions. 
For each luminosity-selected sample, partitioning the range of $1\le \delta(R_{f1})\le 6$ into fifty bins, we count the numbers of the wall 
galaxies whose local density lies in each bin of $\delta(R_{f1})$. Let $n_{1,\alpha},\ n_{2,\alpha},\ n_{3,\alpha},\ n_{4,\alpha}$ denote the 
numbers of the wall galaxies that fall in the $\alpha$th bin, in the sample {\cal I},\ {\cal II},\ {\cal III} and {\cal IV}, respectively. 
Define $n_{{\rm min},\alpha}$ as  $n_{{\rm min},\alpha}\equiv {\rm Min}\{n_{k,\alpha}\}_{k=1,}^{4}$.
We extract $n_{{\rm min},\alpha}$ wall galaxies from each of the four samples at the $\alpha$th bin. Repeating this process for 
all of the $50$ bins, we now have four controlled samples of the wall galaxies which have the identical density distribution. 
 
The density distributions of the wall galaxies belonging to the original and controlled samples are shown as histograms in the top and 
bottom panels of Figure \ref{fig:dendis_hden}, respectively.  As can be seen, while the density distributions from the original four samples 
are different from one another (top panel), the four controlled samples have the identical density distributions, being completely overlapped 
with one another. Taking the same steps but with the field galaxies with $\delta(R_{f1})<1$, we also construct four controlled samples of the 
field galaxies. Figure \ref{fig:dendis_lden} shows the same as Figure \ref{fig:dendis_hden} but for the case of the field galaxies.

The average tidal coherence $\langle\vert\hat{\bf e}(R_{f1})\cdot\hat{\bf e}(R_{f2})\vert\rangle$, is evaluated over the wall and field 
galaxy grids separately as: 
\begin{equation}
\label{eqn:eecor}
\langle\vert\hat{\bf e}(R_{f1})\cdot\hat{\bf e}(R_{f2})\vert\rangle = 
\frac{1}{N_{\rm g}}\sum_{\beta=1}^{N_{g}}\vert\hat{\bf e}(R_{f1})\cdot\hat{\bf e}(R_{f2})\vert_{\beta}\, ,
\end{equation}
where $N_{g}$ is the number of the wall (field) galaxies belonging to each controlled sample and 
$\vert\hat{\bf e}(R_{f1})\cdot\hat{\bf e}(R_{f2})\vert_{\beta}$ is the value of the tidal coherence computed at the $\beta$-th wall (field) 
galaxy grid. The associated errors, $\sigma_{e}$, for each controlled sample are also evaluated by employing the Bootstrap method as
\begin{equation}
\label{eqn:boot}
\sigma^{2}_{e} = 
\frac{1}{N_{\rm boot}}\sum_{\gamma=1}^{N_{\rm boot}}
\left[\langle\vert\hat{\bf e}(R_{f1})\cdot\hat{\bf e}(R_{f2})\vert\rangle_{\gamma}- 
\langle\vert\hat{\bf e}(R_{f1})\cdot\hat{\bf e}(R_{f2})\vert\rangle\right]^{2}\, ,
\end{equation}
where $N_{\rm boot}=1000$ is the number of the bootstrap resamples and 
$\langle\vert\hat{\bf e}(R_{f1})\cdot\hat{\bf e}(R_{f2})\vert\rangle_{\gamma}$ denotes the average tidal coherence obtained 
from the $\gamma$-th Bootstrap resample. 

Figure \ref{fig:eecor_hden} shows the mean values of the tidal coherence and the local densities averaged over the wall galaxies 
belonging to each of the four controlled luminosity-selected samples in the top and bottom panels, respectively.  As can be seen, while the 
mean local densities are indeed well controlled, the four controlled samples yield significantly different values of the tidal coherence from one 
another. 
The sample {\cal IV} yields the highest value of the tidal coherence, while the lowest value is found in the sample {\cal I}.  We note a clear 
trend that the degree of the tidal coherence gradually increases with the decrement of the luminosities of the wall galaxies. 
This result indicates that the least luminous wall galaxies (i.e., sample {\cal IV}), preferentially reside in the regions where the directions 
of the minor eigenvectors of the tidal shear field are coherent over large scales. 

We follow the same procedure but with the field galaxies and show the results in Figure \ref{fig:eecor_lden}. 
Note that the field galaxies also exhibit a similar trend: the least luminous field galaxies preferentially reside in the grids with highest tidal 
coherence.  Unlike the case of the wall galaxies, however, the lowest tidal coherence is exhibited not by the sample {\cal I} but by the 
sample {\cal II}, although the difference in the mean values of the tidal coherence between the samples {\cal I} and {\cal II} is not significant.  

To examine if and how the results change with the smoothing scale, we repeat the same analysis but with using two different 
smoothing scales of $R_{f2}$.  The left (right) panels of Figure \ref{fig:eecor_scale} display the mean tidal coherence of the 
field (wall) galaxies belonging to each of the four controlled samples for the cases of $R_{f2}=20\,h^{-1}$Mpc (top panel) and 
$R_{f1}=10\,h^{-1}$Mpc (bottom panel).  As can be seen, the same trend is found for all of the four cases, which proves the robustness 
of the trend against the variation of the smoothing scale, $R_{f2}$. 

\section{Dependence of the Galaxy Morphology on the Tidal Coherence}\label{sec:eecor_type}

Now, we are going to investigate how the trend between the tidal coherence and the galaxy luminosity varies with the galaxy 
morphology. Let $P_{\rm max}$ denote the maximum value among four probabilities of a given galaxy being classified as 
four morphological types:  
\begin{equation}
\label{eqn:pmax}
P_{\rm max}\equiv {\rm max}\{P({\rm E}),\ P({\rm S0}),\ P({\rm Sab}),\ P({\rm Scd})\}\, .  
\end{equation}
The late-type spirals, early-type spirals, lenticulars and ellipticals are determined as those which distinctively satisfy the conditions of 
$P_{\rm max}=P({\rm Scd})$, $P_{\rm max}=P({\rm Sab})$, $P_{\rm max}=P({\rm S0})$, $P_{\rm max}=P({\rm E})$,  respectively. 

Let us start with the Scd galaxies. Basically we redo the same analysis described in Section \ref{sec:eecor_lum} but only with 
Scd galaxies: (i) make four controlled luminosity-selected subsamples of the Scd galaxies according to their $M_{r}$ values; 
(ii) separate the Scd galaxies into the wall and field category for each controlled subsample.
(iii) evaluate the mean values of the tidal coherence as well as the local density averaged over the Scd wall and field 
galaxies separately for each controlled subsample. Then, we follow the same procedure for the Sab, S0 and E galaxies. 

The top-left, bottom-left, top-right, bottom-right panels of Figure \ref{fig:eecor_type_lum_hden} plot the mean values of the tidal coherence 
averaged over the Scd, Sab, S0 and E wall galaxies, respectively, versus the four controlled subsamples.  As can be seen, 
the Scd, Sab and S0 wall galaxies exhibit the same trend  that we have witnessed in Section \ref{sec:eecor_lum}: the more luminous 
galaxies tend to reside in the regions with lower tidal coherence. 
Meanwhile, E wall galaxies behave differently: the sample {\cal I} containing the most luminous elliptical wall galaxies 
have almost the same tidal coherence as the sample {\cal IV} containing the dimmest counterparts. The lowest average value of the tidal 
coherence is found in the sample {\cal II}. 

\begin{figure}
\begin{center}
\includegraphics[scale=0.95]{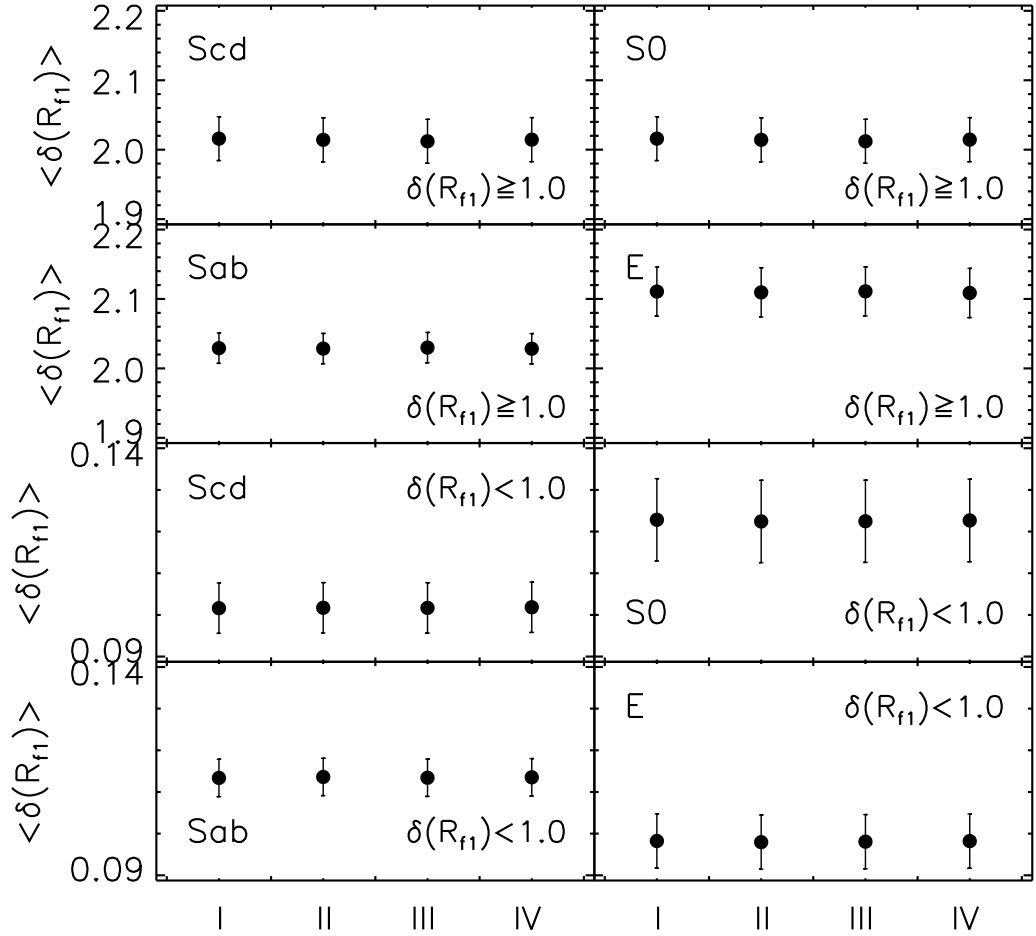}
\caption{Mean values of the local densities for the four different cases of the galaxy morphological type 
in the wall (top two panels) and field (bottom two panels) environments. }
\label{fig:meanden}
\end{center}
\end{figure}

Figure \ref{fig:eecor_type_lum_lden} plots the same as Figure \ref{fig:eecor_type_lum_hden} but for the field galaxies. 
The Scd and S0 field galaxies show the same trend as their wall counter parts. For the Sab field galaxies, however, no strong trend is 
found, which implies that the luminosities of the Sab field galaxies may be independent of the tidal coherence. 
For the elliptical field galaxies, the highest value of the tidal coherence is found in the sample {\cal IV} while the other three samples 
show no difference in the value of the tidal coherence. This result indicates that the growth of only very dim elliptical field galaxies are 
affected by the tidal coherence. Figure \ref{fig:meanden} plots the mean local densities averaged over the wall and field galaxies 
belonging to each of the four controlled samples at fixed morphology, which ensures that the four controlled samples indeed have the 
same density distributions. 

\section{Summary and Discussion}\label{sec:con}

We have investigated the effect of the tidal coherence on the galaxy luminosity by utilizing the density fields constructed by 
\citet{wan-etal12} and the flux-limited galaxy catalog from the SDSS DR10 \citep{tem-etal14}. 
From the catalog,  we have selected the local galaxies within a volume of $\sim 180^{3}\,h^{-3}\,{\rm Mpc}^{3}$ centered on the Milky 
Way and determined the values of the local density field smoothed on the nonlinear scale ($R_{f1}$) at the locations of the selected 
galaxies. 
Depending on the value of the local density ($\delta$), the selected local galaxies are divided into the wall ($\delta\ge 1$) and field 
($\delta<1$) categories. We have also determined two minor eigenvectors of the tidal shear fields smoothed on the nonlinear and 
linear scales ($R_{f2}$) and then computed the tidal coherence as the cosines of the alignment angles between the two minor 
eigenvectors at each galaxy grid.

\begin{figure}
\begin{center}
\includegraphics[scale=0.95]{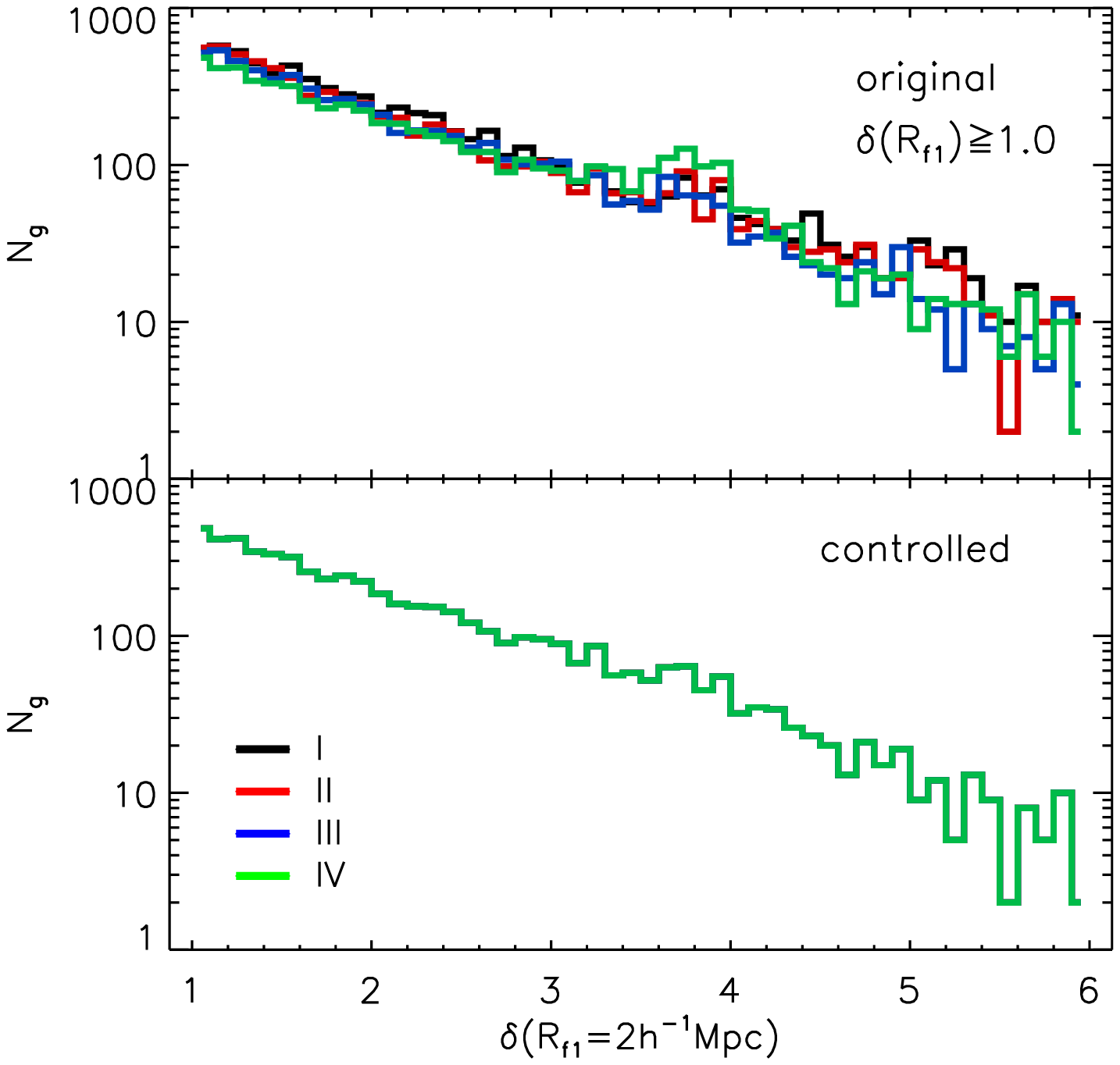}
\caption{Local density distributions of the galaxies with $\delta(R_{f1})\ge 1$ belonging to four uncontrolled 
and four controlled luminosity-selected samples in the top and bottom panels, respectively.}
\label{fig:dendis_hden}
\end{center}
\end{figure}

\begin{figure}
\begin{center}
\includegraphics[scale=0.95]{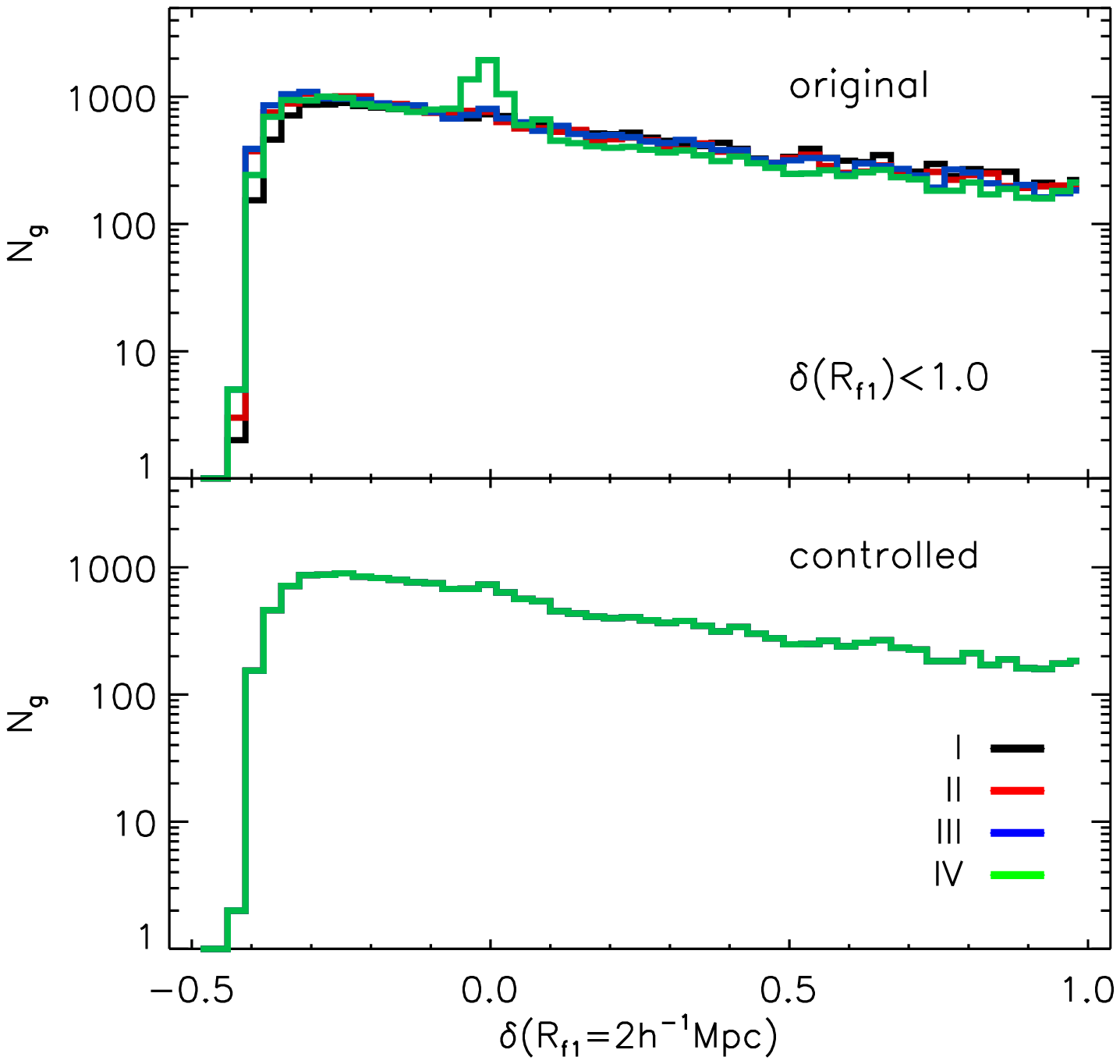}
\caption{Same as Figure \ref{fig:dendis_hden} but for the galaxies with $\delta(R_{f1})<1$.}
\label{fig:dendis_lden}
\end{center}
\end{figure}

\begin{figure}
\begin{center}
\includegraphics[scale=0.95]{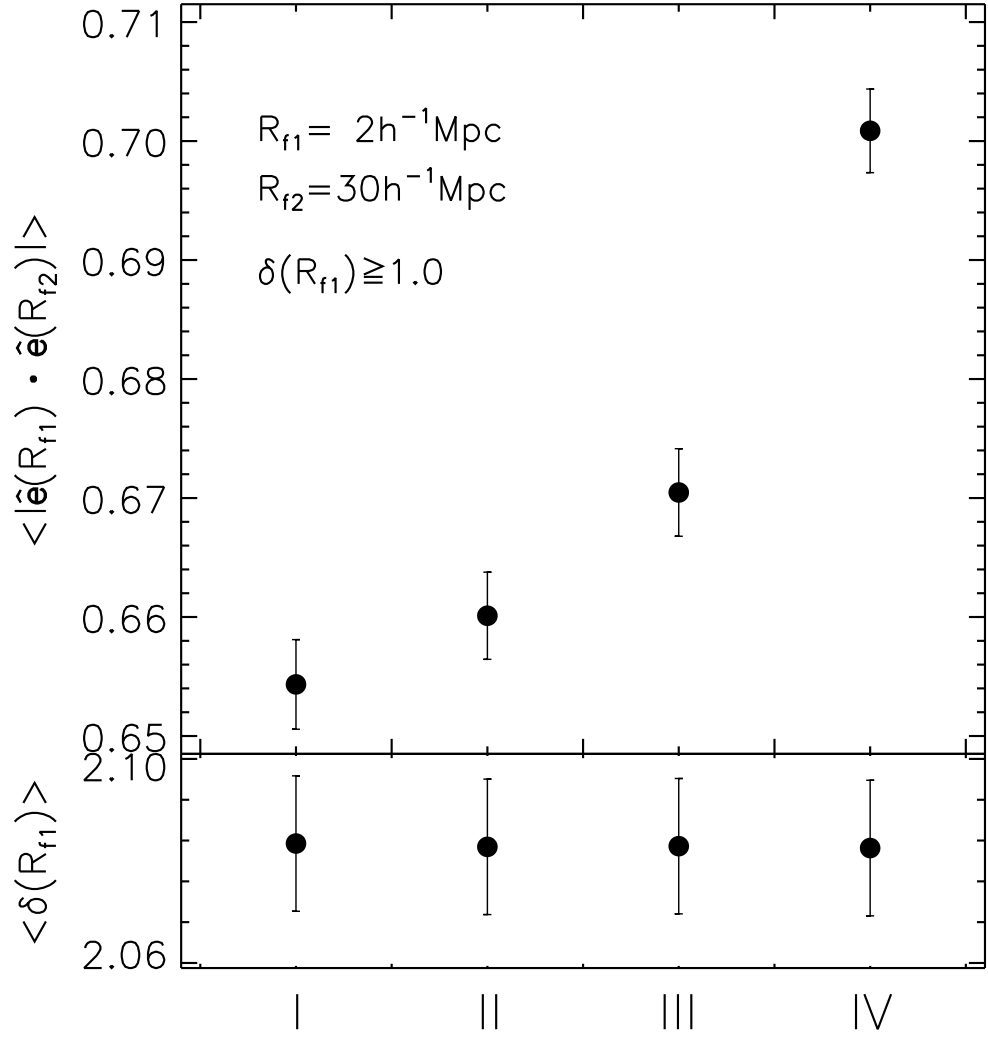}
\caption{Mean values of the tidal coherence (top panel) and the local density (bottom panel) averaged over 
the galaxies with $\delta(R_{f1})\ge 1$ belonging to each of the four controlled luminosity-selected samples}
\label{fig:eecor_hden}
\end{center}
\end{figure}

\begin{figure}
\begin{center}
\includegraphics[scale=0.95]{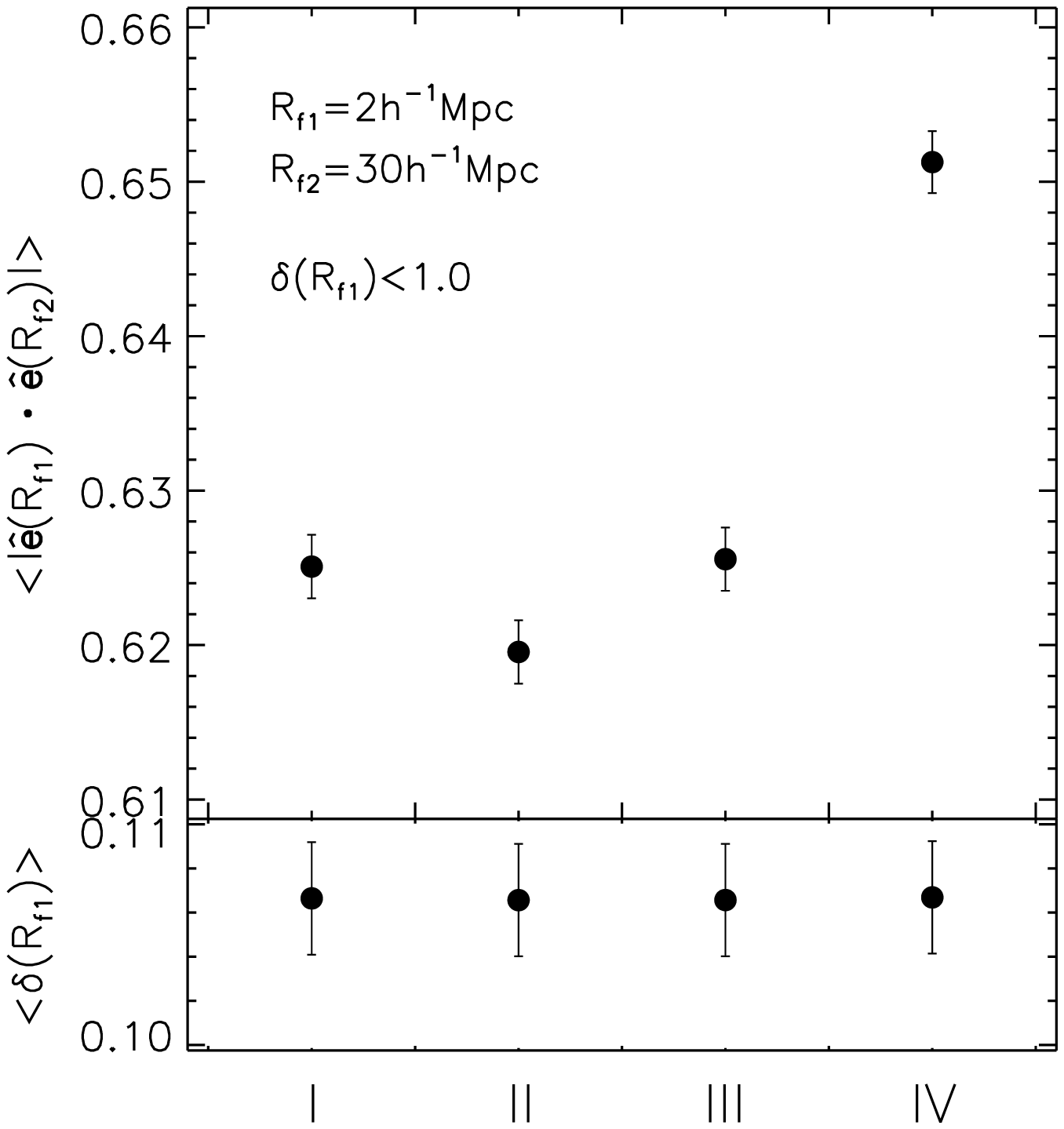}
\caption{Same as Figure \ref{fig:eecor_hden} but for the galaxies with $\delta(R_{f1})<1$.}
\label{fig:eecor_lden}
\end{center}
\end{figure}

\begin{figure}
\begin{center}
\includegraphics[scale=0.95]{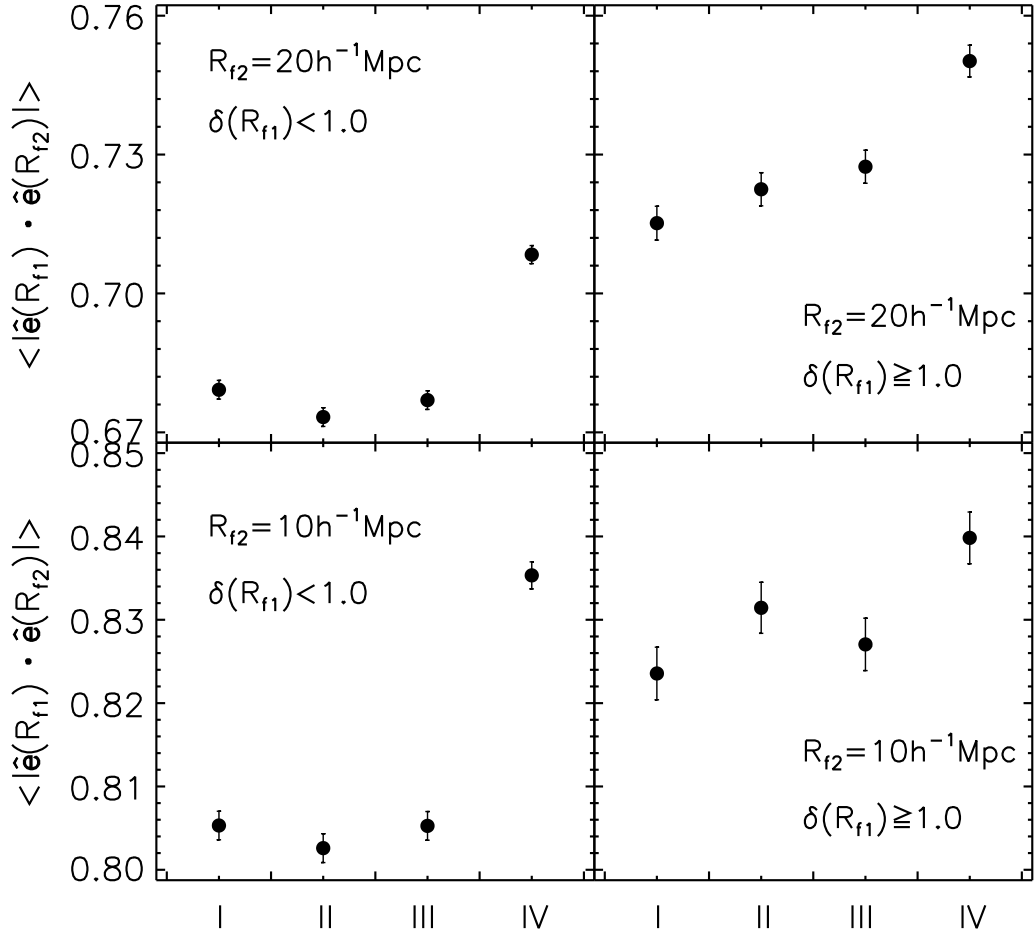}
\caption{Same as Figure \ref{fig:eecor_hden} (left panel) and Figure \ref{fig:eecor_lden} (right panel) 
but for the cases of $R_{f2}=20\,h^{-1}$Mpc and $R_{f1}=10\,h^{-1}$Mpc in the top and bottom panels, 
respectively.}
\label{fig:eecor_scale}
\end{center}
\end{figure}

Four luminosity-selected samples ({\cal I},\ {\cal II},\ {\cal III} and {\cal IV}) of the target galaxies are created from the wall and field 
category separately and controlled to have identical distributions of the local densities (Figures \ref{fig:dendis_hden}-\ref{fig:dendis_lden}). 
Finally, we have evaluated the mean value of the tidal coherence averaged over each of the four controlled samples and found a clear 
trend that the samples containing more luminous galaxies yield lower tidal coherence. This result implies that the local galaxies 
preferentially reside in the regions where the eigenvectors of the tidal fields are coherent in their directions over large-scales, 
whether they belong to the field or wall environments (Figure \ref{fig:eecor_hden}-\ref{fig:eecor_lden}). 
Fixing the nonlinear smoothing scale at $2\,h^{-1}$Mpc throughout the analysis, we have shown that this trend is robust against the 
variation of the linear scales from $10\,h^{-1}$Mpc to $30\,h^{-1}$Mpc (Figure \ref{fig:eecor_scale}).  

Our physical explanation for this trend is as follows. In the wall environment, the filamentary merging and infall of the satellites is the 
main route toward the growth of the galaxies. In a region with high tidal coherence where the minor eigenvectors of the tidal fields on 
largely different scales are well aligned, the filamentary merging between the satellites could not occur frequently 
since most of the satellites would move coherently in the directions perpendicular to the filament axes (i.e., the directions of the 
minor eigenvectors). 
The higher tidal coherence a region has, the more coherent tangential motions the regional satellites would develop over larger scales. 
In other words, since the merging and infall of the satellites into a larger galaxy system can be expedited only when 
the satellites develop radial motions along the filament axes aligned with the minor eigenvectors of the tidal fields, 
the growth of those galaxies located in the regions with high tidal coherence would be severely obstructed.  In consequence, the 
locations of more massive wall galaxies (or more luminous wall galaxies) would be biased toward the regions with lower tidal coherence. 

The same logic can be applied to the field galaxies whose growth is mainly driven by the diffusive accretion of matter and gas particles 
\citep{FM10}. In the regions with higher tidal coherence, the matter and gas particles would also develop tangential motions in the 
plane normal to the minor eigenvectors coherent over large scales. Thus, the growth of the field galaxies located in those regions with 
higher tidal coherences would also be retarded, which would in turn lead to the biased population of the massive field galaxies 
in the regions with lower tidal coherence. 
It is worth mentioning here that the dominant effect on the luminosity of a galaxy is its gravitational interaction with the accreting 
satellites and gas, the rate of which depends on the local densities. Since the main focus of the current analysis is to find an 
{\it independent} effect of the large-scale tidal coherence on the galaxy luminosity, we had to nullify this dominant effect of the 
gravitational interaction by controlling the four samples to have the same distributions of the local densities. 

The detected trend is consistent with the result of \citet{zomg1} from zoom hydrodynamic simulation that the "stalled" galactic halos are 
embedded in a straight bulky filament while the "accreting" halos are located at the junction of multiple narrow filaments. \citet{zomg1} 
explained that the satellite infall and merging into the stalled galactic halos are hindered by their tangential motions perpendicular to the 
bulk filaments while the radial flows of the satellites onto the junction of the multiple narrow filaments lead to an efficient growth of the 
accreting halos \citep[see also][]{zomg3}.
Comparison between their numerical and our observational results indicates that inside a bulk straight filament should correspond to 
a wall region with high tidal coherence while a junction of multiple narrow filaments should be a wall region with low tidal coherence. 

Note that our work in fact provides a more fundamental dichotomy between the "stalled" and "accreting" galaxies. The former should not be 
limited to the galaxies embedded in straight bulk filaments in the wall environment. Rather it should be extended to a broad class of the 
galaxies located in the regions with high tidal coherence, whether it is a wall or a field environment.  Likewise, the latter should not be 
limited to the galaxies located at the junction of multiple narrow filaments. Rather it should include the galaxies located in the regions with 
low tidal coherence. 
In contrast to the claim of the subsequent hydrodynamic work of \citet{zomg3} that no difference exists in the baryonic properties between 
the stalled and accreting halos, our work provides an observational evidence for the existence of difference in the stellar contents 
between the two. 

\begin{figure}
\begin{center}
\includegraphics[scale=0.95]{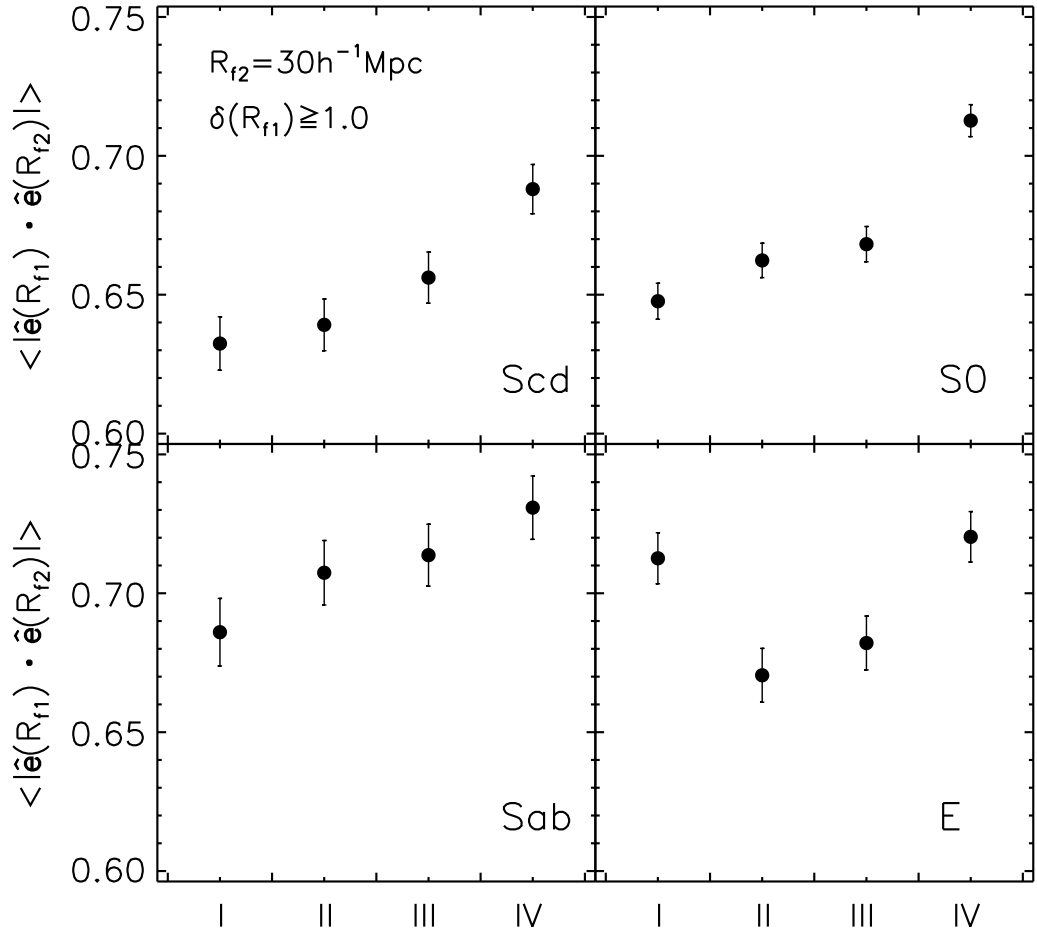}
\caption{Same as Figure \ref{fig:eecor_hden} but with the galaxy morphology fixed at the Scd, 
Sab, S0 and E types in the top-left, bottom-left, top-right and bottom-right panels, respectively.} 
\label{fig:eecor_type_lum_hden}
\end{center}
\end{figure}

\begin{figure}
\begin{center}
\includegraphics[scale=0.95]{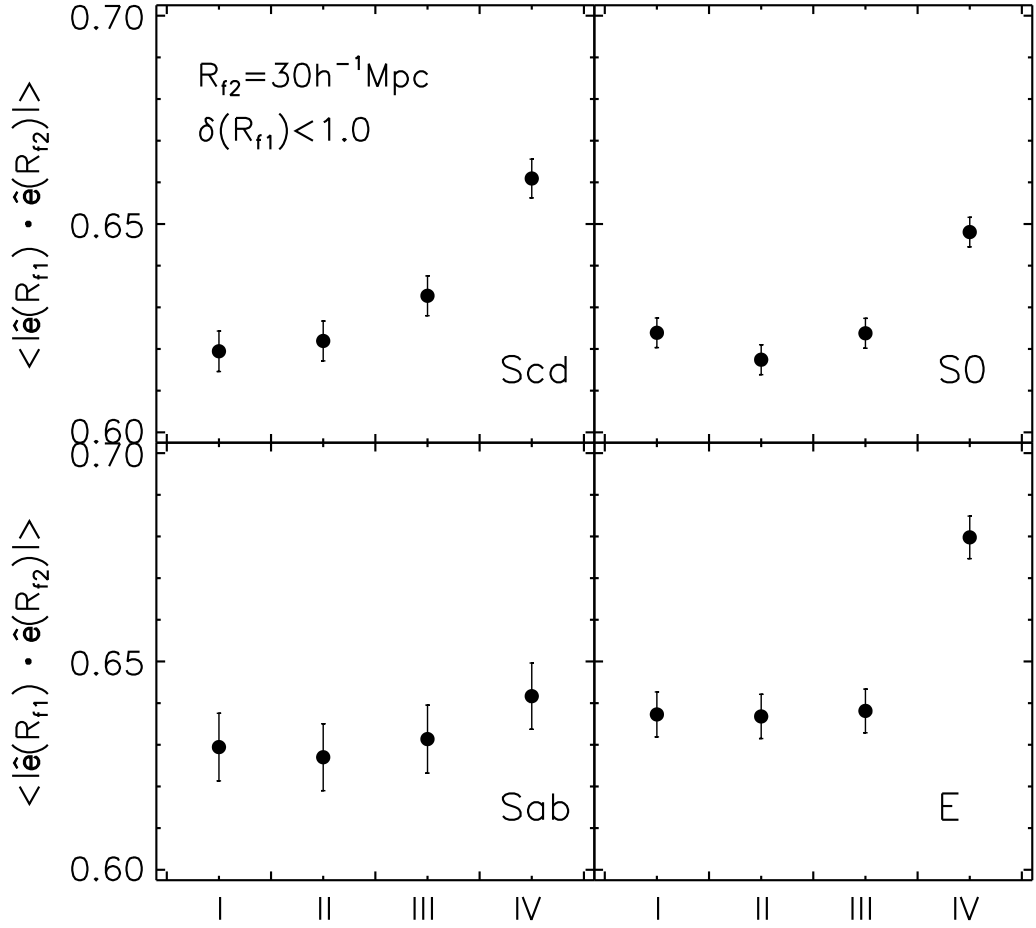}
\caption{Same as Figure \ref{fig:eecor_type_lum_hden} for the case of $\delta(R_{f1})<1$.}
\label{fig:eecor_type_lum_lden}
\end{center}
\end{figure}

We have also examined if and how the trend depends on the galaxy morphology, redoing the same analyses at fixed morphological 
types (i.e., Scd, Sac, S0 and E types, separately). In the wall environments, the same trend has been found from the Scd, Sab and S0 
galaxies: the sample {\cal IV} ({\cal I}) containing the least (most) luminous wall galaxies yields the highest (lowest) tidal coherence. 
Meanwhile, a different trend has been witnessed for the case of the E wall galaxies: the highest values of the tidal coherence has been 
found not only from the sample {\cal IV} containing the dimmest ones but also from the sample {\cal I} having the most luminous 
E wall galaxies (Figure \ref{fig:eecor_type_lum_hden}). 
We suspect that the majority of the sample {\cal I} which has exhibited the high degree of tidal coherence should be the brightest 
cluster galaxies (BCG) located at the centers of the galaxy clusters.  Given that a galaxy cluster is usually located in a thick bulky filament, 
the tidal coherence is expected to have a high value at the location of its center. Once a satellite falls within the virial radius of its host 
cluster, however, they would develop radial motions in the direction toward the BGC. In other words, for the exceptional case of the BGCs, 
the luminosity and the tidal coherence can develop a positive correlation.  
In the field environment, a similar trend, albeit weak, has been detected from the Scd, S0, and E galaxies: the least luminous Scd, S0 
and E field galaxies have been found to reside preferentially in the regions with highest tidal coherence. Whereas, no trend from the 
Sab galaxies has been found (Figure \ref{fig:eecor_type_lum_lden}). This result implies that the tidal coherence does not make any 
significant contribution to the growth of the Sab field galaxies. 

The direction of our future work is to investigate the reason that the early type spiral galaxies located in the field environment shows no 
dependence on the large-scale tidal coherence unlike the other morphological types and also to see if the galaxy properties other than the 
luminosity and morphological types such as the star formation rate (SFR) depend on the tidal coherence. 
Although the SFR of a galaxy was found not to depend sensitively on the large-scale environment \citep{YH16,bey-etal17}, 
this issue is still inconclusive since the previous works focused only on the correlations between the SFR  and the large-scale densities.  
If the SFR be also found to depend on the large-scale tidal coherence, it will support more strongly our claim that the growth of the 
galaxies in the cosmic web is strongly, if not dominantly, affected by the large-scale tidal coherence. 

\acknowledgements

I acknowledge the support of the Basic Science Research Program through the National Research Foundation (NRF) of Korea 
funded by the Ministry of Education (NO. 2016R1D1A1A09918491).  I was also partially supported by a research grant from the 
NRF of Korea to the Center for Galaxy Evolution Research (No.2017R1A5A1070354).

\clearpage
\clearpage
\clearpage
\clearpage
\clearpage
\clearpage
\clearpage
\clearpage
\clearpage
\begin{deluxetable}{cccc}
\tablewidth{0pt}
\setlength{\tabcolsep}{5mm}
\tablecaption{}
\tablehead{sample & $M_{r}$ & $N_{g}$}
\startdata
{\cal I} & $[-23.28, \   -19.95)$  & $29882$ \\
{\cal II} &$[-19.95 , \   -19.15)$  &$29883$ \\
{\cal III} & $[-19.15, \    -18.27)$ &$29883$ \\
{\cal IV} & $[-18.27,  \   -9.95]$  & $29884$
\enddata
\label{tab:mr}
\end{deluxetable}

\end{document}